\date{Rev. 22/II/13 JM}
\title{Computing higher homotopy groups is $W[1]$-hard}
\newif\ifcmts
\cmtstrue

\author{
{\sc Ji\v{r}\'{\i} Matou\v{s}ek}\thanks{Supported
by the  ERC Advanced Grant No.~267165.}
\\
   {\footnotesize Department of Applied Mathematics and}\\[-1.5mm]
   {\footnotesize Institute of Theoretical Computer Science (ITI)}\\[-1.5mm]
   {\footnotesize  Charles University, Malostransk\'{e} n\'{a}m. 25}\\[-1.5mm]
{\footnotesize  118~00~~Praha~1,
   Czech Republic, and}\\%[-1.5mm]
{\footnotesize    Institute of  Theoretical Computer Science}\\[-1.5mm]
{\footnotesize    ETH Zurich,
      8092 Zurich, Switzerland}
}

\documentclass[11pt,a4paper]{article}

\usepackage{amssymb,amsmath,amsthm}
\usepackage{graphicx}
\usepackage{url}
\usepackage{a4wide}

\newtheorem{theorem}{Theorem}%[section]

\newcommand{\heading}[1]{\vspace{1ex}\par\noindent{\bf #1}}

\newcommand{\Z}{{\mathbb{Z}}}
\newcommand{\Q}{{\mathbb{Q}}}

\newcommand\VV{\mathcal{V}}

\newcommand\makevec[1]{{\bf #1}}

\def \vv {\makevec{v}}
\def \ww {\makevec{w}}

\def\bzero{\makevec{0}}

\DeclareMathOperator{\rk}{rk}
\newcommand{\thedim}{k}

\def\:{\colon}

\long\def\onefigure#1#2{%  #1 picture,  #2  caption
\begin{figure*}[tbp]
\begin{center}
#1
\end{center}
\caption{#2}
\end{figure*}
}

\def\immediateFigure#1{%
\smallskip\begin{center}#1\end{center}\smallskip }

\newcommand{\labfig}[2]  % labeled figure
{\onefigure{\mbox{\includegraphics{Figures/#1}}}{\label{f:#1} #2} }

\newcommand{\labfigw}[3]  % labeled figure with prescribed width
{\onefigure{\mbox{\includegraphics[width=#2]{Figures/#1}}}{\label{f:#1} #3}}

\newcommand{\immfig}[1]  % immediate figure
{\immediateFigure{\mbox{\includegraphics{Figures/#1}}}}

\newcommand{\immfigw}[2] % immediate figure with prescribed width
{\immediateFigure{\mbox{\includegraphics[width=#2]{Figures/#1}}}}

\ifcmts
\newcommand{\jirka}[1]{\ifhmode\newline\fi\marrow \textsf{\color{magenta}*** (JIRKA: ) #1\newline}}
\else
\newcommand{\jirka}[1]{\relax}
\fi

\begin{document}

\maketitle

\begin{abstract} Recently it was shown that, for every fixed $k\ge 2$,
given a finite simply connected simplicial complex $X$, the $k$th homotopy
group $\pi_k(X)$ can be computed in time polynomial in the number $n$
of simplices of $X$. We prove that this problem is $W[1]$-hard w.r.t.~the 
parameter $k$ even for $X$ of dimension~$4$, and thus very unlikely to 
admit an algorithm with running time bound $f(k)n^C$ for an absolute 
constant~$C$. We also simplify, by about 20 pages, a 1989 proof by Anick
that, with $k$ part of input, the computation of the rank of $\pi_k(X)$ is 
\#P-hard.
\end{abstract}

\paragraph{Introduction. } The homotopy groups $\pi_k(X)$, $k=1,2,\ldots$,
belong among the most important and most puzzling invariants of
a topological space $X$ (see, e.g., \cite{Ravenel,Kochman} for
the amazing adventure of computing the homotopy groups of
spheres, where only partial results have been obtained in spite
of an enormous effort).\footnote{We recall that 
$\pi_\thedim(X)$  defined as the set of all homotopy classes
of \emph{pointed} continuous maps $f\:S^\thedim\to Y$ (where $S^k$
stands for the $k$-dimensional sphere), i.e.,
maps $f$ that send a distinguished
point $s_0\in S^\thedim$ to a distinguished point $x_0\in X$.
Here two pointed maps $f,g$ are \emph{homotopic} if $f$ can be
continuously deformed into $g$ while keeping the image of $s_0$
fixed; this means that there is a continuous map 
$F\:S^k\times [0,1]\to X$ with $F(\cdot,0)=f$, $F(\cdot,1)=g$,
and $F(s_0,\cdot)=x_0$. 
 Strictly speaking, one should
really write $\pi_\thedim(X,x_0)$ instead of $\pi_k(X)$,
but for a path-connected $X$, the choice of $x_0$ does not matter.
Each $\pi_\thedim(X)$, $k\ge 1$, is a group, which for $\thedim\ge 2$
is Abelian, but the definition of the group operation is not important
for us at the moment.
}

In this note we consider the (theoretical) complexity of computing
$\pi_k(X)$, for given $k$ and $X$. We assume that the space
$X$ is given as a finite simplicial complex, and the size of
the input is measured as the number of simplices of~$X$.

It is well known that the fundamental group $\pi_1(X)$ is
uncomputable, as follows from undecidability of the word
problem in groups \cite{Novikov:UndecidabilityWordProblem-1955}.
On the other hand, given a $1$-connected  $X$,
i.e., one with $\pi_1(X)$ trivial, 
there are algorithms that compute  $\pi_\thedim(X)$, 
for every given~$\thedim\ge 2$ (more precisely, it is known
that for a finite simplicial complex $X$, $\pi_k(X)$ is a
finitely generated Abelian group, and the algorithms compute
its isomorphism type, i.e., express it as a direct sum of cyclic groups).
The first such algorithm is due to Brown
\cite{Brown:FiniteComputabilityPostnikovComplexes-1957},
and newer ones have been obtained as a part of general
computational frameworks in algebraic topology
due to Sch\"on \cite{Schoen-effectivetop} and due to
%, Smith \cite{smith-mstructures}, 
and Sergeraert and his co-workers (e.g.,
 \cite{Sergeraert:ComputabilityProblemAlgebraicTopology-1994,SergerGenova,Real96}). Recently it was shown by \v{C}adek et al.~\cite{polypost} that, for
every fixed $k\ge 2$, $\pi_k(X)$ can be computed in polynomial time,
where the polynomial depends on~$k$.

As for lower bounds, Anick \cite{Anick-homotopyhard} proved that
computing $\pi_\thedim(X)$ is \#P-hard, where $X$ can even be assumed 
to be a $4$-dimensional $1$-connected
space, but, crucially, $\thedim$ is regarded as a 
part of the input. The hardness also applies to the potentially easier
problem of computing only the rank of $\pi_\thedim(X)$,
i.e., the number of direct summands isomorphic to $\Z$.
In Anick's original result, the space $X$ is not given as a simplicial
complex, but in another, considerably more compact representation,
but it was shown by \v{C}adek et al.~\cite{ext-hard} that Anick's 
representation can be converted into a simplicial complex with
only polynomial-time overhead, and thus the \#P-hardness result also
applies to (1-connected, 4-dimensional, finite) simplicial complexes.

\paragraph{Results. } Given that $\pi_k(X)$ is polynomial-time computable
for $k$ fixed, it is natural to ask whether it is \emph{fixed-parameter
tractable}, i.e., computable in time $f(k)n^C$ for some absolute constant $C$
and some function $f$ of $k$; see, e.g., \cite{RolfFPT} for an introduction
to the field of \emph{parameterized complexity}, which considers
this kind of questions. We show that this is very unlikely.

\begin{theorem}\label{t:} The problem of computing $\pi_k(X)$ for
a $4$-dimensional $1$-connected simplicial complex $X$ with $n$ simplices
is $W[1]$-hard (with parameter $k$).
\end{theorem}

We refer to \cite{RolfFPT} for the definition of the class of $W[1]$-hard 
problems. Here it suffices to say that no problem in this class
is known to be fixed-parameter tractable, and it is widely believed 
no $W[1]$-hard problem is fixed-parameter tractable  (this is somewhat
similar to the widely held belief that P$\ne$NP). 

The proof of Theorem~\ref{t:} is very short and simple if we take two
reductions from Anick \cite{Anick-homotopyhard} for granted. At the same time,
it gives a considerable simplification of Anick's \#P-hardness proof,
replacing about 20 pages of Anick's paper and a substantial
part of its technical contents.

\heading{Vests. } Anick \cite{Anick-homotopyhard} defines an auxiliary
computational problem called \emph{vest} (``vector evaluated after
a sequence of transformations''). The input instance $\VV$ of a vest
is given by a (column) vector $\vv\in\Q^d$, a list
 $(T_1,T_2,\ldots,T_m)$ of rational $d\times d$
matrices, and a $h\times d$ rational matrix $S$ (for some natural
numbers $d,m,h$). 
The \emph{$M$-sequence} of such a $\VV$  is the integer sequence
$(M_1,M_2,\ldots)$, 
where
\[
M_k:=|\{(i_1,i_2,\ldots,i_k): S T_{i_k}T_{i_{k-1}}\cdots T_{i_1}\vv =\bzero\}|,
\]
with $\bzero$ denoting the (column) vector of $h$ zeros.

Anick \cite{Anick-homotopyhard} makes a connection of vests
to the ranks of homotopy groups, which relies on other papers
and apparently is not easy to trace down in detail. First,
given an instance $\VV$ of a vest, one can construct a (suitable finite 
presentation
of) a certain algebraic structure called a \emph{123H-algebra} $A$,
such that a suitable integer sequence associated with $A$ 
(the \emph{Tor-sequence} of $A$) equals the $M$-sequence of $\VV$.
This is stated as \cite[Thm.~3.4]{Anick-homotopyhard}, but the proof
refers to \cite[Thm.~1.3]{anick1985diophantine}, which expresses the
desired connection in a different language, and it is actually a
special case of Theorem~7.6 of~\cite{anick-generic}. 

Second, given the considered presentation of $A$, 
one can construct a $4$-dimensional
$1$-connected cell complex $X$ (which, in turn, can be converted
into a simplicial complex in view of \cite{ext-hard}) such that
the Tor-sequence of $A$ and the sequence of ranks $(\rk \pi_2(X),
\rk\pi_3(X),\ldots)$ are \emph{rationally related}, which in particular
means that the first $k$ terms of one of the sequences can be computed
from the first $O(k)$ terms of the other sequence, in polynomial time
(with a \emph{polynomial} dependence on $k$ as well). 
The construction of $X$ from $A$ relies on Roos~\cite{roos-relations}. 

It would be nice to streamline these reductions and have them
summarized at one place, but here we take them for granted.
In particular, they imply that $W[1]$-hardness or \#P-hardness of the
vest problem implies $W[1]$-hardness or \#P-hardness of the problem
of homotopy group computation considered in Theorem~\ref{t:}, respectively.

\begin{proof}[\bf Hardness of vests: proof of Theorem~\ref{t:}.]
Given a graph $G$ on $n$ vertices, the problem of testing the
existence of a clique (complete subgraph) on $k$ vertices in $G$
is one of the most famous and useful $W[1]$-complete problems \cite{RolfFPT}.

For a given $G$ and $k$, we construct a vest $\VV=(\vv,T_1,\ldots,T_m,S)$ 
for which, with $s=k+{k\choose 2}$,
the $s$th term of the $M$-sequence is 
$M_{s}=s! C_k$, where $C_k$ is the number of $k$-cliques in~$G$.

Let us call a vector $\ww$ a \emph{current vector} if it has the form
$T_{i_j}T_{i_{j-1}}\cdots T_1\vv$ for some $j$ and some $i_1,\ldots,i_j$.
We will not describe the $T_i$ explicitly as matrices; rather, we will
say how $T_i$ transforms the current vector $\ww$ into $T_i\ww$,
where we assume that all the components of $\ww$ that are not
explicitly mentioned in such a description are left unchanged by~$T_i$.

The initial vector $\vv$, and thus all current vectors,
have $d=n+2m+1$ components (where $m$, yet unspecified, is the number
of the $T_i$). The first $n$ components, called the \emph{vertex components},
are in one-to-one correspondence with the vertices of $G$.
 Then there is a special component that equals $1$ in $\vv$, as well as in all
current vectors (thus, no $T_i$ is going to change it); all the other
components of $\vv$ are set to $0$. Finally, for each $T_i$, we have
two \emph{private components} in each current vector, 
which are changed by that $T_i$ but by no other $T_j$.

Let $a$ and $b$ be the two private components belonging to some $T_i$;
we let $T_i$ transform them to $a+1$ and $b+a$, respectively
(note that the $1$ in $a+1$ really means adding the special component to $a$).
This guarantees that after at most one application of $T_i$, the second private
component of $T_i$ is $0$, while two or more applications 
of $T_i$ make it nonzero.

The $T_i$ in $\VV$ are actually indexed by $V(G)\cup E(G)$, vertices and edges
of the given graph (thus, $m=n+|E(G)|$). The $T_i$ corresponding
to a vertex $v$ increments the vertex entry of $v$ in the current vector
by $k-1$, while the $T_i$ corresponding to an edge $\{u,v\}$ decrements
the vertex entries of $u$ and $v$ each by~$1$.

It remains to specify the matrix $S$. We construct it as
 a zero-one matrix with a single
$1$ per row. Thus, the effect of multiplying the current vector by 
$S$ is selecting
certain components, and we construct $S$ so that exactly the vertex components
and the second private component of each $T_i$ are selected; therefore,
$h=n+m$.

Because of the private components, the vector
$ST_{i_s}T_{i_{s-1}}\cdots T_1\vv$ can be
zero only if $i_1,\ldots,i_s$ are all distinct. 
Then it is easy to argue that exactly
$k$ vertex $T_i$'s and $k\choose 2$ edge $T_i$'s must be used, corresponding
to the vertex set and edge set of a $k$-clique in $G$, respectively. 
Since the ordering of such $T_i$'s is arbitrary, each $k$-clique contributes
$s!$ to~$M_s$.
\end{proof}

The problem of counting $k$-cliques in a given graph, with $k$ a part
of input, is \#P-complete, and thus the above proof also
provides the promised simplification of Anick's \#P-hardness proof.

\bibliographystyle{alpha}
\bibliography{Postnikov.bib}

\end{document}